# Structural Decomposition of Moran's Index by Getis-Ord's Indices

Yanguang Chen

(Department of Geography, College of Urban and Environmental Sciences, Peking University, 100871, Beijing, China. Email: chenyg@pku.edu.cn)

**Abstract**: Moran's index and Getis-Ord's indices are important statistical measures of spatial autocorrelation analysis. Each of them has its own function and scope of application. However, the association of Moran index with Getis-Ord index is not clear. This paper is devoted to deriving and verify the relationships between Moran's index and Getis-Ord's indices using mathematical reasoning and empirical analysis. Getis-Ord's indices are employed to decompose Moran's index. The results show that there is a strict nonlinear relationship between Moran's index and Getis-Ord's indices. Moran's index consists of four components: global Getis-Ord's index, sum of local Getis-Ord's indices, number of elements, and size correlation function. Thus the mathematical structure of Moran's index is revealed. A theoretical discovery is that the characteristics of spatial autocorrelation depends on the relationship between the global Getis-Ord's index and the sum of local Getis-Ord's indices, as well as the number of spatial elements. Local Getis-Ord's indices proved to be equivalent to the potential indices based on gravity model. A conclusion can be drawn that Moran's index is related to gravity model, and thus spatial autocorrelation is associated with spatial interaction. This indicates that weak spatial interaction leads to not significant spatial autocorrelation. The conclusion is supported by observational data. This study not only helps to better understand the basic statistics of spatial analysis, but also contributes to the further development of spatial autocorrelation theory.

# 1 Introduction

Many statistic measures can be used in spatial autocorrelation analysis, and different measures have different statistical significance and application scope. Among various spatial autocorrelation measures, Moran's index is the basic and most important measure (Cliff and Ord, 1981; Griffith, 2003; Haggett *et al*, 1977; Moran, 1950; Odland, 1988), while Getis-Ord's index is a concise and practical measurement (Getis and Ord, 1992). Measurements are the bricks of theory, models are the buildings of theory, and the relationships between measurements and the relationships between models as well as the relationships between models and measurements are the roads and bridges of theory construction. In order to develop the theory and method of geospatial analysis, it is necessary to reveal the relationships between various statistics and spatial analysis models. However, so far, the mathematical and numerical relationships between Moran's index and Getis-Ord's index has not been fully revealed. This is not only unfavorable for students to deeply understand the similarities and differences between Moran's index and Getis-Ord's index, but also unfavorable for the needs of theoretical development of spatial autocorrelation.

A discovery is that Moran's index can be decomposed using Getis-Ord's indices. The results contribute to a deeper understanding of the Moran's index and Getis-Ord's indices, enabling better utilization of them for spatial statistical analysis. Based on standardized variables and normalized spatial weight matrix, both Moran's index and Getis-Ord's index can be expressed in quadratic form. The quadratic forms provides convenience for revealing the mathematical relationship between the two spatial statistics. As long as the mathematical relationship is clear, the numerical relationship will be clear. This research is devoted to deriving the strict relationships between Moran's index and Getis-Ord's index. The derivation results will show the intrinsic ingredients of Moran's index. The rest parts of this paper will be organized as follows. In Section 2, the mathematical relation between Moran's index and Getis-Ord's index will be derived step by step, and the mathematical structure of Moran's index will be revealed. In Section 3, the cities in Beijing, Tianjin and Hebei region, China, are taken as example for empirical analysis to verify the results of mathematical derivation. In Section 4, several related questions are discussed, and finally, in Section 5, the discussion will be concluded by summarizing the main points of this work.

# 2 Theoretical results

## 2.1 Basic concepts and expressions

For the sake of simplicity and intuitiveness in the reasoning process, it may be helpful to do some preparing work on concepts and expressions. Firstly, let's see the quadratic form expression of the Getis-Ord's index. Suppose that there are $N$ spatial elements in a geographical region. The size of each element is $x_i$ or $x_j$, and distance between element $i$ and element $j$ is $r_{ij}$ ($i, j$ =1, 2, …, $N$). Normalized spatial weight can be expressed as below:

$$w_{ij} = \frac{v_{ij}}{\sum_{i=1}^{N}\sum_{j=1}^{N} v_{ij}}, \tag{1}$$

where $v_{ij}=f(r_{ij})$ denotes spatial contiguity between places $i$ and $j$, and $w_{ij}$ refers to globally normalized weight. What is called global normalization is the normalization method based on sum. Its property is that each weight value comes between 0 and 1, and the sum of all the weight values is equal to 1. Size measure can be normalized as follows

$$p_i = \frac{x_i}{\sum_{i=1}^{N} x_i}, \quad p_j = \frac{x_j}{\sum_{j=1}^{N} x_j}, \tag{2}$$

where $p_i$ and $p_j$ represent globally normalized $x_i$ and $x_j$. Thus, the global Getis-Ord's index can be simplified as a quadratic form as below (Chen, 2020)

$$G = \sum_{i=1}^{N}\sum_{j=1}^{N} w_{ij} p_i p_j = \mathbf{p}^{\mathrm{T}} \mathbf{W} \mathbf{p}. \tag{3}$$

This is actually a spatial correlation function, that is, a correlation function based on spatial displacement. Accordingly, the local Getis-Ord's index of the $i$th element can be computed by

$$G_i = \sum_{j=1}^{N} w_{ij} p_i = \mathbf{W} \mathbf{p}. \tag{4}$$

Suppose that $G_L$ denotes the sum of local Getis-Ord's indices. The result is

$$G_L = \sum_{i=1}^{N} G_i = \sum_{i=1}^{N}\sum_{j=1}^{N} w_{ij} p_i = \mathbf{o}^{\mathrm{T}} \mathbf{W} \mathbf{p}. \tag{5}$$

where $\mathbf{o}$ is an ones vector, i.e., $\mathbf{o}=[1, 1,…,1]^{\mathrm{T}}$. Further, a simple size correlation function can be expressed as

$$C_f = \sum_{i=1}^{N} p_i^2 = \mathbf{p}^T \mathbf{p}, \tag{6}$$

which is the inner product of the probability measure vector, **p**. This correlation function describes size correlation of elements in a region. Based on equation (6), a correlation dimension can be defined for multifractals (Feder, 1988; Vicsek, 1989).

## 2.2 Relation between Moran's index and Getis-Ord's indices

Now, we can derive the relation between Getis-Ord's indices and Moran's index. In this work, Getis-Ord's indices include global Getis-Ord's index and local Getis-Ord's indices. However, Moran's index only involves global Moran's statistics. Based on normalized size variable, Moran's index can be expressed as

$$I = \frac{\sum_{i=1}^{N}\sum_{j=1}^{N} v_{ij}(p_i - \bar{p})(p_j - \bar{p})}{\sum_{i=1}^{N}\sum_{j=1}^{N} v_{ij} \cdot \frac{1}{N}\sum_{i=1}^{N}(p_i - \bar{p})^2} = \frac{\sum_{i=1}^{N}\sum_{j=1}^{N} v_{ij}(p_i - \frac{1}{N})(p_j - \frac{1}{N})}{\sum_{i=1}^{N}\sum_{j=1}^{N} v_{ij} \cdot \frac{1}{N}\sum_{i=1}^{N}(p_i - \frac{1}{N})^2}. \tag{7}$$

Equivalently, equation (7) can be transformed into the following form

$$I = \sum_{i=1}^{N}\sum_{j=1}^{N} w_{ij} \frac{(Np_i - 1)(Np_j - 1)}{\frac{1}{N}\sum_{i=1}^{N}(Np_i - 1)^2}. \tag{8}$$

which can be converted to a quadratic form (Chen, 2023), that is

$$I = \sum_{i=1}^{N}\sum_{j=1}^{N} w_{ij} \frac{Np_i - 1}{\sqrt{\frac{1}{N}\sum_{i=1}^{N}(Np_i - 1)^2}} \frac{Np_j - 1}{\sqrt{\frac{1}{N}\sum_{j=1}^{N}(Np_j - 1)^2}} = \sum_{i=1}^{N}\sum_{j=1}^{N} w_{ij} z_i z_j = \mathbf{z}^T \mathbf{W} \mathbf{z}, \tag{9}$$

where $\mathbf{W}=[w_{ij}]$ refers to globally normalized spatial weight matrix, and **z** is standardized size vector based on $z$-score. Developing the denominator of equation (8) yields

$$\frac{1}{N}\sum_{i=1}^{N}(Np_i - 1)^2 = \frac{1}{N}\sum_{i=1}^{N}(N^2 p_i^2 - 2Np_i + 1) = (N\sum_{i=1}^{N} p_i^2 - 2\sum_{i=1}^{N} p_i + 1). \tag{10}$$

Considering the property of probability measure and the definition of correlation function based on probability, we have

$$\frac{1}{N}\sum_{i=1}^{N}(Np_i - 1)^2 = N\sum_{i=1}^{N} p_i^2 - 1 = NC_f - 1. \tag{11}$$

Substituting equation (11) into equation (8) yields

$$I = \sum_{i=1}^{N}\sum_{j=1}^{N} w_{ij} \frac{N^2 p_i p_j - Np_i - Np_j + 1}{NC_f - 1}. \tag{12}$$

Thus we have

$$I = \frac{1}{NC_f - 1}(N^2 \sum_{i=1}^{N}\sum_{j=1}^{N} w_{ij} p_i p_j - 2N \sum_{i=1}^{N}\sum_{j=1}^{N} w_{ij} p_i + \sum_{i=1}^{N}\sum_{j=1}^{N} w_{ij}). \tag{13}$$

Based on matrix and vector, equation (13) can be equivalently expressed as

$$I = \frac{1}{N\mathbf{p}^T\mathbf{p} - 1}[N^2 \mathbf{p}^T \mathbf{W}\mathbf{p} - 2N\mathbf{o}^T \mathbf{W}\mathbf{p} + 1]. \tag{14}$$

From equation (14) it follows

$$I = \frac{1}{NC_f - 1}(N^2 G - 2N \sum_{i=1}^{N} G_i + 1) = \frac{1}{NC_f - 1}(GN^2 - 2G_L N + 1), \tag{15}$$

which can be easily verified by observed data. Equation (15) gives the relation between Moran's index and Getis-Ord's indices. A normalized global Getis-Ord's index can be defined based on equation (3) and equation (6) as follows

$$g = \frac{G}{C_f} = \frac{\mathbf{p}^T \mathbf{W} \mathbf{p}}{\mathbf{p}^T \mathbf{p}}, \tag{16}$$

where $g$ denotes normalized global Getis-Ord's index. Equation (16) is a form of Rayleigh quotient (Chen, 2023; Xu, 2021). Thus, equation (15) can be expressed as

$$I = \frac{1}{N - 1/C_f}(gN^2 - 2g_L N + \frac{1}{C_f}), \tag{17}$$

where

$$g = \frac{G_L}{C_f} = \frac{1}{C_f} \sum_{i=1}^{N}\sum_{j=1}^{N} w_{ij} p_i = \frac{\mathbf{o}^T \mathbf{W} \mathbf{p}}{\mathbf{p}^T \mathbf{p}}, \tag{18}$$

which can be treated as the sum of standardized local Getis-Ord's indices.

As indicated above, Moran's index is the most important measurement of spatial autocorrelation. Now, we can analyze the mathematical structure and composition of Moran's index. According to equation (15), Moran's index contains four components, namely, global Getis-Ord's index, the sum of local Getis-Ord's indices, size correlation function, and number of spatial elements (Table 1). Specifically, the composition is as follows. (1) Global Getis-Ord's index. The global Getis-Ord's index determines the positive spatial autocorrelation information contained in Moran's index. Since the value of global Getis-Ord's index is greater than 0, the higher Getis-Ord's index, the higher Moran's index, and it changes in the positive direction. In the physical sense, it can be proved that the global Getis-Ord's index is equal to the sum of gravity. (2) Local Getis-Ord's indices. The higher the sum of the local Getis-Ord's index, the smaller Moran's index, and it changes in the negative direction. In the physical sense, it was proved that the local Getis-Ord's indices are equivalent to the potential indices, which can be derived from gravity model. (3) Size correlation function. Size correlation function is the correlation function without spatial lag. Formally, the correlation function is the inner product of the normalized size vector. In the physical sense, it is the second moment of the probability distribution. For given spatial scale, the size correlation function is a constant. (4) The number of elements. For the research of urban systems, the element number is the number of cities in a geographical region. It reflects the macro state number of a system. The larger the number of macro states, the higher the spatial entropy, and the more complex the system may be.

Table 1 Structural decomposition of Moran's index and mathematical or statistical meaning of different components

| Composition | Symbol | Physical meaning | Mathematical relation | Statistic effect |
| --- | --- | --- | --- | --- |
| Global Getis-Ord's index | $G$ | Gravity | Linear relation | Determining the positive spatial autocorrelation information contained in Moran's $I$ |
| Local Getis-Ord's indices | $G_L$ | Potential | Linear relation | Determining the negative spatial autocorrelation information contained in Moran's $I$ |
| Size correlation | $C_f$ | Moment | Inverse relation | Inversely proportional to the absolute value of Moran's $I$ |
| Number of elements | $N$ | State | Nonlinear relation | The relationship between the element number and Moran's $I$ is complex |

## 2.3 Structural analysis of Moran's index

Exploring the conditions of significant spatial autocorrelation is theoretically significant for the development of spatial analysis methods. For a spatial population, or for a very large spatial sample, the special condition of no spatial autocorrelation is $I=0$ in theory, thus, according to equation (15),

we have

$$GN^2 - 2G_L N + 1 = 0, \quad (19)$$

which represents the quadratic equation of element number based on Getis-Ord's index. Solving this quadratic equation yields

$$N_{1,2} = \frac{2G_L \pm \sqrt{4G_L^2 - 4G}}{2G} = \frac{G_L \pm \sqrt{G_L^2 - G}}{G}. \quad (20)$$

If two roots are expressed separately, then we have

$$N_1 = \frac{G_L - \sqrt{G_L^2 - G}}{G}, \quad (21)$$

$$N_2 = \frac{G_L + \sqrt{G_L^2 - G}}{G}. \quad (22)$$

This means that for a given number of geographical elements, spatial autocorrelation may disappear under certain conditions.

For real spatial systems, there may be three situations and four autocorrelation patterns as follows. (1) For $G_L^2 > G$, equation (19) has two real roots. In this case, if the element number $n$ is less than $N_1$ or greater than $N_2$, the spatial autocorrelation is positive; if the element number $n$ comes between $N_1$ and $N_2$, the spatial autocorrelation is negative. (2) For $G_L^2 = G$, equation (19) has unique real root. In this case, there is no spatial autocorrelation or positive spatial autocorrelation. This situation is not easy to encounter in the real world. (3) For $G_L^2 < G$, equation (19) has no real root. In this case, the spatial autocorrelation is generally positive. Based on the above analysis, it can be concluded that positive spatial autocorrelation may have multiple reasons, while negative spatial autocorrelation has only one reason (Table 2).

**Table 2 Spatial autocorrelation types determined by the relationship between global Getis-Ord's index and local Getis-Ord's indices as well as element number**

| Index relation | Quadratic equation root | Element number | Spatial autocorrelation |
|---|---|---|---|
| $G_L^2 > G$ | Two real roots | $N < N_1$, or $N > N_2$ | Positive spatial autocorrelation |
| | | $N_1 < N < N_2$ | Negative spatial autocorrelation |

| $G_L^2=G$ | One real root | $N = N_1 = N_2$ | Zero or positive spatial autocorrelation |
| --- | --- | --- | --- |
| $G_L^2<G$ | No real root | No relation of $N$ to $N_1$ and $N_2$ | Positive spatial autocorrelation |

**Note**: The judgment results in the table can be understood based on the knowledge of inequalities in elementary mathematics. The symbol $N$ represent observed spatial element number in the real world.

However, for a smaller spatial sample, the precondition of no significant spatial autocorrelation is different. In this case, the critical value is $I=1/(1-N)$. Therefore, we can derive a relation from equation (15) as below:

$$I - \frac{1}{1-N} = \frac{1}{NC_f - 1}(N^2G - 2NG_L + 1) - \frac{1}{1-N} = 0. \tag{23}$$

From equation (23) it follows

$$-GN^3 + 2(G + 2G_L)N^2 - (C_f + 2G_L + 1)N + 2 = 0. \tag{24}$$

By solving this cubic equation, we can obtain the special condition that the spatial samples have no significant autocorrelation. The roots of the equations will give the numbers of elements with spatial autocorrelation at a special critical values.

## 3 Empirical analysis

### 3.1 Study area and data

The correctness of mathematical reasoning results can be verified through case analysis. The observed data of urban systems in the real world can be used to confirm the derived relations above. The study area is a specific region consisting of one province (Hebei) and two municipalities directly under the central government (Beijing and Tianjin), which is called the Beijing-Tianjin-Hebei (BTH) region of China (Long and Chen, 2019). The region is also termed Jing-Jin-Ji (JJJ) area in literature (Hu, 2023; Lu, 2015). There are 13 cities at prefecture level and above and 35 cities at county level and above in the study area. Three data sources are as follows. The spatial distances are measured by traffic mileage, and the data were extracted by ArcGIS from digital maps of BTH region. City sizes can be measured by city population and urban nighttime light (NTL) area (Table 3). The data of NTL area are defined within built-up area of cities in BTH region (Chen and Long, 2021; Long and Chen, 2019). The data of city population come from China Population Census in 2000 (the fifth census) and 2010 (the sixth census). The spatial proximity is defined by inverse distance function $v_{ij}=1/r_{ij}$, where $r_{ij}$ denotes the traffic mileage between city $i$ and city $j$. The spatial contiguity matrix can be expressed as $\mathbf{V}=[v_{ij}]=[1/r_{ij}]$, in which the diagonal elements are defined as zero. As indicated above, normalizing the spatial contiguity matrix $\mathbf{V}$ by sum yields a standard spatial weight matrix $\mathbf{W}=[w_{ij}]$. As a globally normalized weight matrix, the summation of its elements equals 1.

Table 3 The measures and data sources for testifying the theoretically derived relations

| Measure | Symbol | Meaning | Data source | Year |
| --- | --- | --- | --- | --- |
| **Distance** | $r_{ij}$ | Interurban distance | Extraction by ArcGIS | 2010 |
| **City size 1** | $x_i$ | City population | The fifth and sixth census of China | 2000, 2010 |
| **City size 2** | $y_i$ | Nighttime light (NTL) area | American NOAA National Centers for Environmental Information (NCEI) | 2000, 2010 |

**Note**: NTL data were processed by Long and Chen (2019) and Chen and Long (2021).

## 3.2 Calculation results

The following data analysis is used to verify theoretical relationships rather than to study real-world problems. We can examine the cities in the study area from two levels. *First of all, the prefecture level cities and cities above prefecture level in the study area are investigated*. For the BTH region, the number of cities at prefecture level and above is $N=13$. Although small samples are not conducive to verifying mathematical laws, they are beneficial for demonstrating calculation methods and testifying mathematical derivation results. Take the area of nighttime lights (NTL area) in BTH region in 2010 as an example to verify the main relations derived above. The directly calculated value of Moran's index is

$$I = \mathbf{z}^T\mathbf{W}\mathbf{z} = -0.1018, \qquad (25)$$

which is the result calculated using the formula of Moran's index. The value of size correlation function is

$$C_f = \mathbf{p}^T\mathbf{p} = 0.2870. \qquad (26)$$

The value of global Getis-Ord's index is

$$G = \mathbf{p}^T\mathbf{W}\mathbf{p} = 0.0070 . \tag{27}$$

The sum of the local Getis-Ord's index is

$$G_L = \mathbf{o}^T\mathbf{W}\mathbf{p} = 0.0944 . \tag{28}$$

Substituting the value of $N$, $C_f$, $G$, and $G_L$ into equation (15) yields

$$\begin{aligned}I &= \frac{1}{NC_f - 1}(GN^2 - 2G_L N + 1)\\ &= \frac{1}{13*0.2870 - 1}(0.0070*13^2 - 2*0.0944*13 + 1) = -0.1018\end{aligned}, \tag{29}$$

which is the result of using the relationship transformation derived above. It should be noted that although the calculation results given here are rounded to 4 decimal places, the actual calculation process is based on 15 decimal places (File S1). Comparing equation (29) with equation (25) shows the value of Moran's index ($I$=-0.1018) converted from global and local Getis-Ord's indices through equation (15) is exactly the same as the Moran's index value ($I$=-0.1018) calculated directly. This implies that the mathematical derivation process is completely correct.

Further, let's verify the quadratic equation of element number based on Getis-Ord's index as well as related inequality. Substituting the value of $G$ and $G_L$ into equation (20) yields

$$N = \frac{G_L \mp \sqrt{G_L^2 - G}}{G} = \frac{0.0944 \mp \sqrt{0.0944^2 - 0.0070}}{0.0070} . \tag{30}$$

Thus we have $N_1$=7.2158 and $N_2$=19.9038. The number of cities, $N$=13, in the study area come between $N_1$ and $N_2$. On the other hand, $G_L^2$=0.0089 > $G$=0.0070. According to the autocorrelation criteria displayed in Table 2, the spatial autocorrelation based on NTL area in 2010 is negative. This is completely consistent with the information reflected by the calculated value of Moran's index, $I$=--0.1018. The other results are tabulated as below (Table 4).

Table 4 The spatial parameters and statistics of the prefecture level and above cities in Beijing, Tianjin, and Hebei region based on city population and NTL area

| Type | Parameter | 2000 | | 2010 | |
| --- | --- | --- | --- | --- | --- |
| | | Population | NTL Area | Population | NTL Area |
| **Calculation** | Moran's $I$ (Direct*) | -0.1513 | -0.1115 | -0.1394 | -0.1018 |
| | $p$-value | 0.1379 | 0.2906 | 0.1769 | 0.3409 |
| **Verification** | $N$ | 13 | 13 | 13 | 13 |
| | $C_f$ | 0.2323 | 0.2791 | 0.2706 | 0.2870 |
|  | $G$ | 0.0063 | 0.0068 | 0.0064 | 0.0070 |
|  | $G_L$ | 0.0913 | 0.0941 | 0.0933 | 0.0944 |
|  | Moran's $I$ (Indirect**) | -0.1513 | -0.1115 | -0.1394 | -0.1018 |
|  | $g$ | 0.0272 | 0.0244 | 0.0235 | 0.0243 |
| **Special** | $N_1$ | 7.3413 | 7.1880 | 7.0590 | 7.2158 |
| **values** | $N_2$ | 21.5446 | 20.3874 | 22.2981 | 19.9038 |
| **Index relation** |  | $G_L{}^2 > G$ | $G_L{}^2 > G$ | $G_L{}^2 > G$ | $G_L{}^2 > G$ |
| **Element number characteristics** |  | $N_1 < N < N_2$ | $N_1 < N < N_2$ | $N_1 < N < N_2$ | $N_1 < N < N_2$ |
| **Spatial autocorrelation** |  | Negative | Negative | Negative | Negative |

**Note**: *This is the direct calculation result by using the formula of Moran's index. **This is the indirect calculation result by using the parameter equation derived in this paper.

The dataset based on prefecture level cities and above is small. As mentioned earlier, there are only 13 elements in total. Although there is no problem in verifying the theoretical relationship with the help of small samples, readers who get accustomed to statistical analysis and are not familiar with the theoretical derivation principle may still question it. If the cities at and above the county level in the study area are considered, there are 35 elements in total ($N$=35). To provide a stable probability distribution, a minimum of 30 sampling points are required. So, generally speaking, 30 observations are the bottom line for statistical analysis (Crilly, 2007). Using the dataset of 35 cities at and above the county level to verify the theoretical relationship derived above, we can see that the calculation results are completely consistent with the theoretical expectation (Table 5). For example, Based on the NTL area data in 2010, the directly calculated value of Moran's index is $I$=0.0145. The value of size correlation function is $C_f$=0.1941. The value of global Getis-Ord's index is $G$=0.00064. The sum of the local Getis-Ord's index is $G_L$=0.0242. Substituting the value of $N$, $C_f$, $G$, and $G_L$ into equation (15) yields $I = (0.00064*35{\wedge}2 - 2*0.0242*35 + 1)/(35*0.1941 - 1) = 0.0145$. By the way, the actual calculation process is based on 15 decimal places rather than 4 decimal places (File S2). Due to $G_L{}^2$=0.00059 < $G$=0.00064, the quadratic equation of element number based on Getis-Ord's index has no real roots. In this case, according to the criteria in Table 2, the spatial autocorrelation is positive, and this consistent with the calculated value of Moran's index.

**Table 5 The spatial parameter and statistics of the county level and above cities in Beijing, Tianjin, and Hebei region based on city population and NTL area**

| Type | Parameter | 2000 | | 2010 | |
|---|---|---|---|---|---|
|  |  | Population | NTL Area | Population | NTL Area |

| Calculation | Moran's $I$ (Direct*) | -0.0121 | 0.0061 | -0.0109 | 0.0145 |
|---|---|---|---|---|---|
| | $p$-value | 0.6394 | 0.8302 | 0.6667 | 0.6345 |
| Verification | $N$ | 35 | 35 | 35 | 35 |
| | $C_f$ | 0.1864 | 0.1955 | 0.2258 | 0.1941 |
| | $G$ | 0.00057 | 0.00061 | 0.00055 | 0.00064 |
| | $G_L$ | 0.0252 | 0.0245 | 0.0250 | 0.0242 |
| | Moran's $I$ (Indirect**) | -0.0121 | 0.0061 | -0.0109 | 0.0145 |
| | $g$ | 0.0030 | 0.0031 | 0.0024 | 0.0033 |
| Special values | $N_1$ | 30.0265 | -- | 29.7252 | -- |
| | $N_2$ | 58.6836 | -- | 60.8676 | -- |
| Index relation | | $G_L^2>G$ | $G_L^2<G$ | $G_L^2>G$ | $G_L^2<G$ |
| Element number characteristics | | $N_1<N<N_2$ | -- | $N_1<N<N_2$ | -- |
| Spatial autocorrelation | | Negative | Positive | Negative | Positive |

**Note**: *This is the direct calculation results by using the formula of Moran's index. **This is the indirect calculation results by using the parameter equation derived in this paper. For the dataset of NTL area, $G_L^2$-$G$<0, so equation (19) has no solution.

In short, the calculation processes based on two sets of data, i.e., the dataset of 13 cities and that of 35 cities, are exactly the same with each other, but the results are not entirely consistent with one another. Based on the above calculation results, the following characteristics of spatial correlation in the Beijing-Tianjin-Hebei urban system can be analyzed. *First, at the larger urban level, i.e., prefecture level cities and above, it manifests as weak negative spatial autocorrelation processes*. This implies a spatial negative feedback process. Whether measured by population sizes or by the NTL areas, the local spatial correlations based on Getis-Ord's index are relatively stronger than the global correlations ($G_L^2>G$). *Second, at the overall level of cities, i.e., county level cities and above, it exhibits both weak spatial positive autocorrelation and weak spatial negative autocorrelation*. The property of spatial autocorrelation depends on the size measures. The spatial autocorrelation based on population size is negative, while the spatial autocorrelation based on NTL area is positive. The population distribution shows a spatial negative feedback process, while the night light distribution shows a spatial positive feedback process. In urban research, NTL data is often used as an alternative measure of urban population size (Long and Chen, 2019). The above spatial autocorrelation analysis implies that using urban NTL data to replace urban population data requires caution. *Third, the urban spatial correlation is mainly based on local effects, and the global spatial correlation is not strong*. This is especially true in terms of population size. This also suggests that there is still a lot of basic work to be done for China's Beijing Tianjin Hebei integration project.

# 4 Discussion

The intrinsic relationship between Moran's index and Getis-Ord's indices has been derived through mathematical reasoning. This is a type of linear and nonlinear interweaving relationship, which can be reflected by equation (15). The result shows that the global Moran's index includes the following information. First, the global Getis-Ord's index, $G$; Second, the sum of local Getis-Ord's indices, $G_L$. Third, the number of elements, $N$. Fourth, size correlation function, $C_f$. The calculated values based on observational data are completely consistent with the theoretical expected values. This means that equation (15) is a strict mathematical relationship, which can provide accurate numerical relationships. Under certain conditions, the local Getis-Ord's indices are equivalent to the potential indices, and the potential indices come from gravity model. On the other hand, the global Getis-Ord's index can be derived from the gravity model. This suggests that Moran's index is associated with gravity model. In this sense, the spatial autocorrelation process is associated with the spatial interaction process.

The above contents mainly involve the relationship between the global Moran's index and Getis-Ord's index. Local Moran's indices are significant for geographical spatial analysis (Anselin, 1995; Anselin, 1996). What is the relationship between the local Moran's indices and Getis-Ord's indices? In fact, the global index relationship suggests the local index relationship. Based on globally normalized spatial weight matrix and $z$-score standardized size vector, the relation between global Moran's index and local Moran's index is

$$I = \sum_{i=1}^{N} I_i, \quad (31)$$

where $I_i$ denote the local Moran's index of the $i$th element. Substituting equation (15) into equation (31) yields

$$\sum_{i=1}^{N} I_i = \frac{1}{NC_f - 1}(N^2 G - 2N \sum_{i=1}^{N} G_i + 1), \quad (32)$$

which shows the relationships between local Moran's index and both global and local Getis-Ord's indices. In spatial autocorrelation research, equation (32) may not be as important as equation (15).

The relation between Moran's index and Getis-Ord's indices can be extended to Geary's coefficient. Geary's coefficient is also an important spatial statistic for geographical analysis (Anselin, 1995; Geary, 1954). Moran's index is based on population standard deviation, while Geary's coefficient is based on sample standard deviation (Chen, 2013). It can be proved that the mathematical relationships between Geary's coefficient and Moran's index is as follows (Chen, 2013)

$$C = \frac{n-1}{n}(\mathbf{o}^\mathrm{T}\mathbf{W}\mathbf{z}^2 - I), \tag{33}$$

where $C$ denotes Geary's coefficient, $\mathbf{o}=[1\ 1\ \ldots\ 1]^\mathrm{T}$ is a one vector, and $\mathbf{z}^2=D_{(z)}\mathbf{z}=[z_1^2\ z_2^2\ \ldots\ z_n^2]^\mathrm{T}$. Here $D_{(z)}$ is the diagonal matrix consisting of the elements of $\mathbf{z}$. Substituting equation (15) into equation (33) yields the relationship between Geary's coefficient and Getis-Ord's indices as below:

$$C = \frac{n-1}{n}[\mathbf{o}^\mathrm{T}\mathbf{W}\mathbf{z}^2 - \frac{1}{NC_f - 1}(GN^2 - 2G_L N + 1)], \tag{34}$$

which can be testified by the observed data of the cities of Beijing, Tianjin, and Hebei region. This suggests that different spatial autocorrelation measures are actually related to one another. Revealing this type of relationships helps to expand and improve the theoretical framework of spatial autocorrelation analysis.

The size correlation function may be confusing. In fact, the above size correlation function is based on characteristic scale. Generally speaking, it is a coefficient rather than function. If variable scales are used instead of characteristic scales, the correlation coefficient will really become a correlation function. Based on a scaling process, correlation function can relate fractal dimension and spatial autocorrelation coefficient. In multifractal theory, correlation dimension can be defined as follows

$$D_2 = \lim_{\varepsilon \to 0} \frac{\ln \sum_{i=1}^{N(\varepsilon)} P_i(\varepsilon)^2}{\ln \varepsilon} \to -\frac{\ln C(\varepsilon)}{\ln \varepsilon}, \tag{35}$$

where $D_2$ denotes correlation dimension, $\varepsilon$ refers to the linear size of fractal units (the side length of the boxes is equivalent to a yardstick), $P_i$ is the ratio of the number of fractal elements appearing in the $i$th unit, $N_i(\varepsilon)$, to the number of all fractal elements, $N(\varepsilon)$, that is, $P_i=N_i(\varepsilon)/N(\varepsilon)$, $C(\varepsilon)$ is the correlation coefficient based on linear size $\varepsilon$. In this case, equation (16) can be expressed as

$$C_f(\varepsilon) = \sum_{i=1}^{N} p_i(\varepsilon)^2 = \mathbf{p}(\varepsilon)^\mathrm{T}\mathbf{p}(\varepsilon), \tag{36}$$

which represents a typical correlation function. It is hard to make it clear in a few lines of words,

and the relevant issues will be specifically explored and discussed in the future.

It should be noted that the Getis-Ord's index may have different forms in the specific calculation process. In the quadratic form expression of Getis-Ord's index given above, the diagonal elements in the outer product matrix of the size vector are taken into account. Corresponding to spatial weight matrix, the diagonal elements of the outer product matrix of the size vector can be omitted. If the diagonal elements are ignored, the expression of Getis-Ord's index can be revised as follows

$$G^* = \frac{1}{1-\mathbf{p}^T\mathbf{p}}\mathbf{p}^T\mathbf{W}\mathbf{p} = \frac{1}{1-\mathbf{p}^T\mathbf{p}}G. \tag{37}$$

Then, the relation between Moran's index and Getis-Ord's index is as below:

$$I = \frac{1}{NC_f - 1}((1-\mathbf{p}^T\mathbf{p})G^*N^2 - 2G_L N + 1). \tag{38}$$

In the previous related equations, using $G^*$ to replace $G$, we have a full set of new expressions. However, it can be seen that the mathematical expression of the new relationship has not changed in essence. In other words, minor changes in form do not affect research conclusions.

No similar studies have been reported in the literature. In terms of empirical analysis, many scholars use spatial autocorrelation analysis methods to study social and economic issues in the Beijing-Tianjin-Hebei region (Chen *et al*, 2016; Lv and Zhang, 2023; Zhou *et al*, 2023). However, in terms of theory and analytical methods, there are not many studies exploring the mathematical relationships between different spatial autocorrelation measures. The novelty of this study is that the mathematical structure and physical meaning of Moran's index are revealed by using Getis-Ord's index. In particular, the quadratic equation of element number based on Getis-Ord's index. The core expressions of this work include equation (15) and equation (19). The shortcomings lie in two aspects. First, no breakthrough in conventional spatial statistical analysis. This study assumes that there are characteristic scales in spatial relationships. In this case, the size correlation function is in fact a size correlation index. In the real world, spatial processes often have no characteristic scale, and thus both Moran's index and Getis-Ord's index are variables instead of constant statistics. However, scale-free spatial distributions and patterns have not been taken into consideration in this research. Second, unsatisfactory case analysis. The spatial autocorrelation of cities in Beijing-Tianjin-Hebei region is not significant. Case analysis is based on the process of insignificant spatial autocorrelation. If there are significant spatial autocorrelation cases as a reference, the empirical analysis is relatively satisfactory. The selection of spatial weight matrix has a certain impact on the conclusions of spatial autocorrelation analysis (Chen, 2012; Getis, 2009; Meng *et al*, 2005). This issue is not covered for the time being since the weight matrix structure does not affect the expression of the measurement relationship, and it is necessary to have a discussion in the future.

## 5 Conclusions

Although Moran's index and Getis-Ord's index represent different types of spatial autocorrelation measures, there is an internal relationships between them. The derivation of this relationship is helpful to reveal the mathematical structure and statistical components of Moran's index. By understanding this structure, we can better conduct spatial autocorrelation analysis and explore geographic spatial processes. Based on the mathematical derivation and empirical analysis, the main conclusions can be reached as follows. *First, there is a strict mathematical relationship between Moran's index and Getis-Ord's indices*. Linear proportional relationship means numerical equivalence, while nonlinear relationship means mathematical equivalence but not numerical equivalence. Moan's index and Getis-Ord's indices have a linear and nonlinear interwoven relationship, which overall exhibits a nonlinear relationship. In this sense, Moran's index is mathematically equivalent to Getis-Ord's index. However, they are not numerically equivalent to one another. *Second, the strength of spatial autocorrelation relies heavily on spatial interactions in geographical processes.* The statistical information of Moran's index can be decomposed into four aspects. Global Moran's index is determined by global Getis-Ord's index, local Getis-Ord's index, size correlation function, and number of elements. Since the global Moran's index is equal to the sum of the local Moran's indices, the statistical information of the local Moran's index can also be decomposed into the above four aspects. Local Getis-Ord's index is equivalent to potential index based on gravity. This suggests that spatial interaction determine the significance of spatial autocorrelation to a degree. *Third, the property of spatial autocorrelation depends on the relationship between the global Getis-Ord's index and the local Getis-Ord's indices, as well as the number of spatial elements*. If there are too few elements, the value of spatial autocorrelation coefficient may be abnormal. For some special number of elements, the spatial autocorrelation is not significant. This may suggest that the role of degrees of freedom in spatial autocorrelation significance testing may be much more complex than previously understood. If the square of the sum of local Getis-Ord's indices is greater than the global Getis-Ord's index, and the number of elements falls between the two real roots of the quadratic equation of the number of elements, the spatial autocorrelation is negative. Otherwise, spatial autocorrelation is positive. Concretely speaking, if the square of the sum of local Getis-Ord's indices is less than the global Getis-Ord's index, we have positive spatial autocorrelation; if the square of the sum of local Getis-Ord's indices is greater than the global Getis-Ord's index, but the number of elements fails to fall between the two real roots of the quadratic equation of the number of elements, we also have positive spatial autocorrelation. Specially, if the square of the sum of local Getis-Ord's indices is equal to the global Getis-Ord's index, we have zero spatial autocorrelation or positive spatial autocorrelation. However, the last scenario rarely occurs in reality.

**Acknowledgement**

This research was sponsored by the National Natural Science Foundation of China (Grant No. 42171192). The support is gratefully acknowledged.